\documentclass[12pt,final,notitlepage,oneside,onecolumn,nobibnotes,
    nofootinbib, superscriptaddress,noshowpacs,centertags]{revtex4}

\usepackage{amsfonts}
\usepackage{amsmath}
\usepackage{amsthm}
\usepackage{amssymb}
\usepackage{graphicx}

\newcommand{\eq}[1]{\begin{equation} #1 \end{equation}}

\newcommand{\eql}[2]{\begin{equation} \label{e:#1} #2 \end{equation}}

\newcommand{\re}[1]{(\ref{e:#1})}

\renewcommand{\i}{\infty}

\begin{document}

\title{ON THE NATURE OF THE CHROMOSPHERE-CORONA TRANSITION REGION OF THE SOLAR ATMOSPHERE}

\author{\copyright\,\, 2010 \quad
        \firstname{O.}~\surname{Ptitsyna}}
      \email{olq1543@gmail.com}
\affiliation{Sternberg Astronomical Institute, Moscow}
\affiliation{Institute for Theoretical and Experimental Physics, Moscow}

\author{\firstname{B.}~\surname{Somov}}
\email{somov@sai.msu.ru}
\affiliation{Sternberg Astronomical Institute, Moscow}
%
%
\begin{abstract}

The distribution of temperature and emission measure in the stationary heated solar
atmosphere was obtained for the limiting cases of slow and fast heating, when either the gas
pressure
or the concentration are constant throughout the layer depth. Under these conditions the temperature
distribution with depth is determined by radiation loss and thermal conductivity.
It is shown that both in the case of slow heating and of impulsive heating, temperatures are
distributed in such a way that \textit{classical collisional heat conduction} is valid in the
chromosphere-corona transition region of the solar atmosphere.

\end{abstract}

\noindent
\hfill ITEP/TH-09/10

\vspace{3mm}

\maketitle

\vspace{3mm}


\vspace{3mm}


%
%

\section{Introduction}

Recently appeared a number of papers \cite{besp-08},\cite{besp-09} claiming that the
chromosphere-corona
transition region of the solar atmosphere should be considered in the non-collisional
approximation.
In these papers it is said that ion-acoustic turbulence is the reason of differentiation of solar
plasma into two regions with high
($ T_{\rm e} \stackrel{ > }{ _{\sim} } 10^6 $~K) and low ($ T_{\rm e} \stackrel{ < }{ _{\sim} } 10^4
$~K) electron temperature, correspondingly.

In this paper we present the solution of the equation of balance between
the thermal heating and radiative cooling with classical electron conductivity. This solution
explain differentiation of solar plasma to the high and low electron temperature. The characteristic
thickness of the chromosphere-corona transition region is greater then thickness corresponding to the
free path for thermal electron collisions and such temperature distribution agrees with observed
solar radiation.

%
%
\section{ Calculation of temperature distribution with depth }

In this paragraph we obtain temperature $T$ distribution with depth $\xi$ for the heating solar
plasma by thermal flux, taking radiation loss into account.

\begin{figure}[ht]
   \includegraphics[scale=0.7]{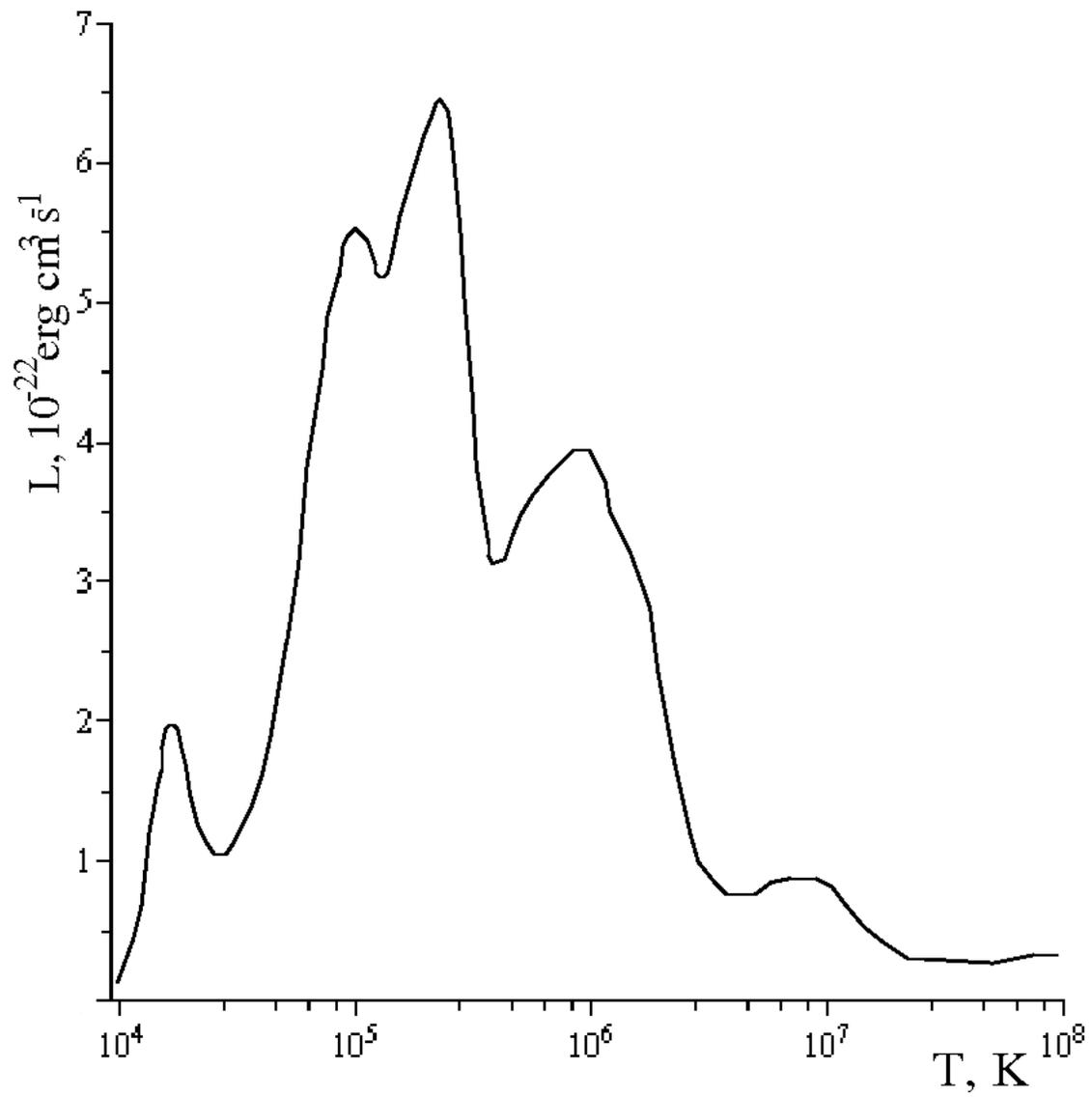}
   \caption{Dependence of the total radiative power loss on the tempetature $L=L(T)$.}
   \label{fig1}
\end{figure}

{\em The thickness\/} is defined as

\eq{
\xi=\int_0^x n(x) dx,
}
where $n(x)$ is the number density of particles.
The thickness represents the number of atoms in the column of unit cross-section along the
direction of thermal flux.
The distribution $T=T(\xi)$  can be obtained with the help of the equation of balance between
the
thermal heating and radiative cooling:

\eql{tepl}{n\frac{d}{d\xi} \left( \kappa n \frac{dT}{d\xi}
\right) = L(T) n^2 - P_\infty,}
where $\kappa$ is electron thermal conductivity,

\eql{kappa}{
\kappa \approx \kappa_0\ T^{5/2}=\frac{1.84 \times 10^{-5} \, T^{5/2}}{ \ln \Lambda} \, , }
\cite{Braginskij},\cite{spitzer},
$L=L(T)$ is the distribution of radiative energy losses with temperature
(see Fig.\ref{fig1}),
à $P=L(T_\infty)n^2$ -- stationary thermal heating at the infinity.

Parameter $L$ of the equation \re{tepl} are known:
$L=L(T)$ was taken from \cite{po4ta}.
To determine the number density dependence  on the temperature $n=n(T)$ let us consider the cases
of slow and fast heating.

One has to specify doundary conditions for the temperature in the equation \re{tepl}:
\eq{\frac{dT}{d\xi}\Big|_{\xi \rightarrow \infty}=0, \quad
T\Big|_{\xi=0} =T_0.}

According to \cite{Shm}, 
to obtain the dependence $T=T(\xi)$
from the equation of the thermal balance \re{tepl} one has to multiply both parts of the
equation \re{tepl} by
$\kappa\frac{dT}{d\xi}$.
Then after simple manipulations we have the following expression for the thermal flux $F$:

\eql{F(T)}{F= \left( \int_{T_\infty}^T 2(L(T')-L(T_{\infty}))
     \kappa n^2 dT' \right)^{\! 1/2} ,}
where the thermal flux $F$ is defined as:
\eql{defF}{F=-\kappa n \frac{dT}{d\xi}.}

To express \re{F(T)} in dimensionless form we multiply \re{F(T)} by

$$\frac{\xi_\i^2}{n_\i^2 \kappa_\i T_\i L(T_\i)},$$
where
$\xi_\i,n_\i,T_\i,\kappa_\i$ are the units of the depth, concentration,
temperature and electron conductivity, correspondingly. 

We obtain
\eq{F= \left( \int_1^T 2 K_1 (L(T)-1)(\frac{T^{5/2}}{ln\Lambda})n^2 dT
       \right)^{\! 1/2} ,}
where $\ln \Lambda$ is the Coulomb logarithm,
\begin{displaymath}
     \ln \Lambda = \left\{
                         \begin{array}{c}
\ln \left( 1.24 \times 10^4\ (T\ T_\infty)^{3/2}/(n\ n_\infty)^{1/2}\right) 
    ,\ \ \ T_e < 5.8\times 10^5\ K,\\
\ln \left( 9.44 \times 10^6\ (T\ T_\infty)/(n\ n_\infty)^{1/2} \right) 
    ,\ \ \ T_e \ge 5.8\times 10^5\ K
                         \end{array}
		    \right.
\end{displaymath}
And $K_1$ 
\eq{K_1=\frac{\xi_\infty^2 L(T_{\infty})}{\kappa_\infty T_\infty}.}
Let us choose the units to be the following (recall that these values are simply our choosen units
and not the actual values at the infinity):
$$T_\infty=10^4 K,$$
$$n_\infty=10^{10} cm^{-3},$$
\eq{ \xi_\infty = n_\infty l_\infty = 3.15 \times 10^{15}\, {\rm cm}^{-2} , \quad
       l_\infty = \left(
       \frac{\kappa_\infty T_\infty }{ L (T_{\infty}) n_\infty^2}
       \right)^{1/2} = 3.15 \times 10^5\, {\rm cm} .}

From equation \re{defF} we obtain the following expression for $\xi$:
\eq{\xi=\int_T^{T_c}\kappa n \frac{dT}{F(T)},}
We choose the origin of $\xi$-coordinate to be the point where the temperature has a fixed value
$T_c$, $T_c=10^6 K$ ( $\xi=0$ when $T=10^6 K$).
\newline

Finally, to get the distribution $T(\xi)$ we can find $F(T)$ from equation \re{F(T)}
 and then, after putting $F(T)$ into \re{defF}, obtain $\xi(T)$ and thus $T(\xi)$.

In the case of fast heating, when concentration is constant throughout the layer depth ($n=const$),
let us set $n=n_\i$, or in dimensionless form $n=1$.

In the case of slow heating gas pressure is constant throughout the layer depth ($p=const$). 
In this case,
because
\eq{p=nk_BT,}
concentration becomes \eq{n=\frac{p}{k_B T}} or in dimensionless form:
\eq{n=\frac{1}{T}.}
\newline

\begin{figure}[ht]
   \includegraphics[scale=0.7]{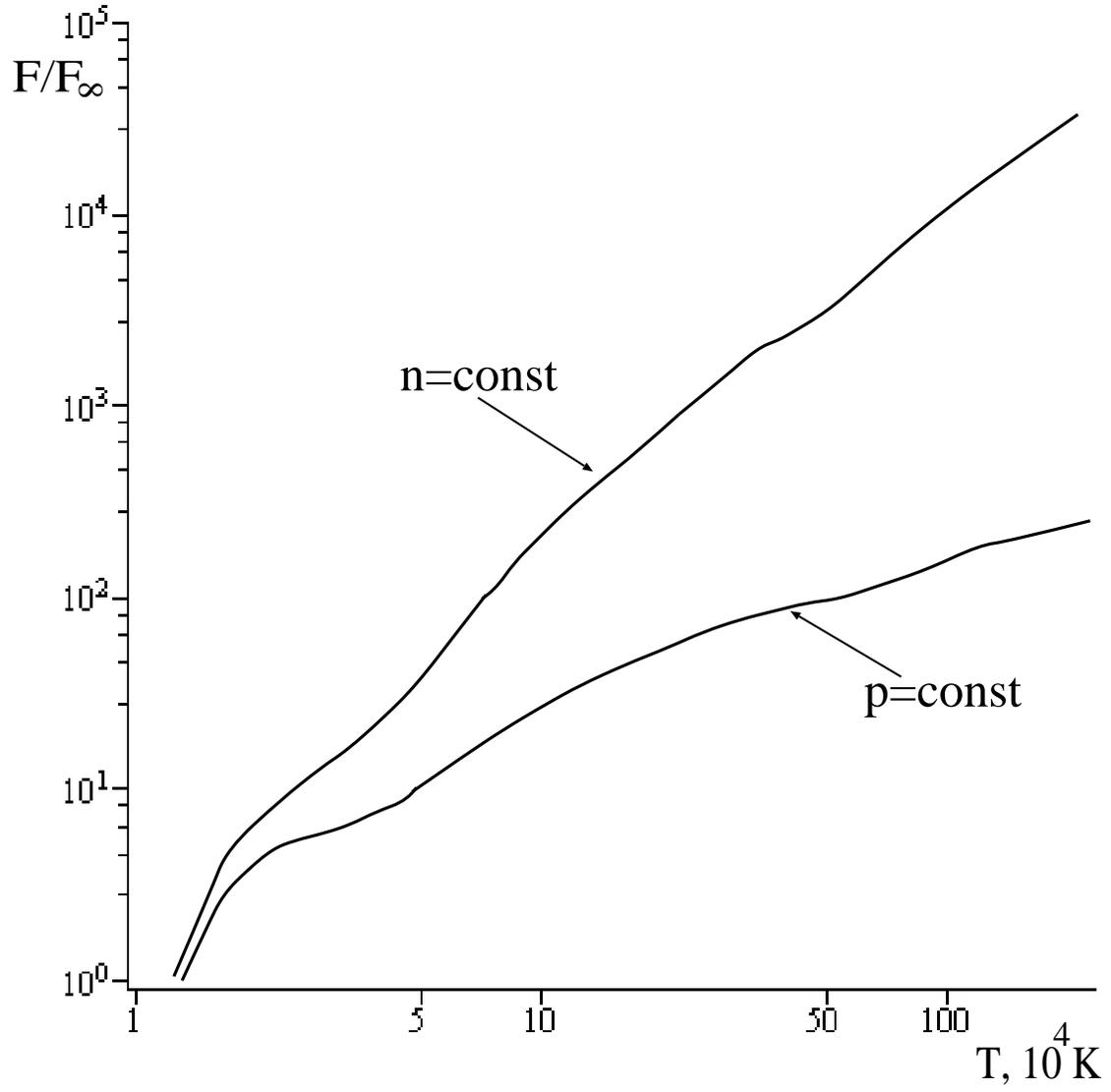}
   \caption{Dependence of the thermal flux on the temperature $F=F(T)$.}
   \label{fig2}
\end{figure}

\begin{figure}[ht]
   \includegraphics[scale=0.7]{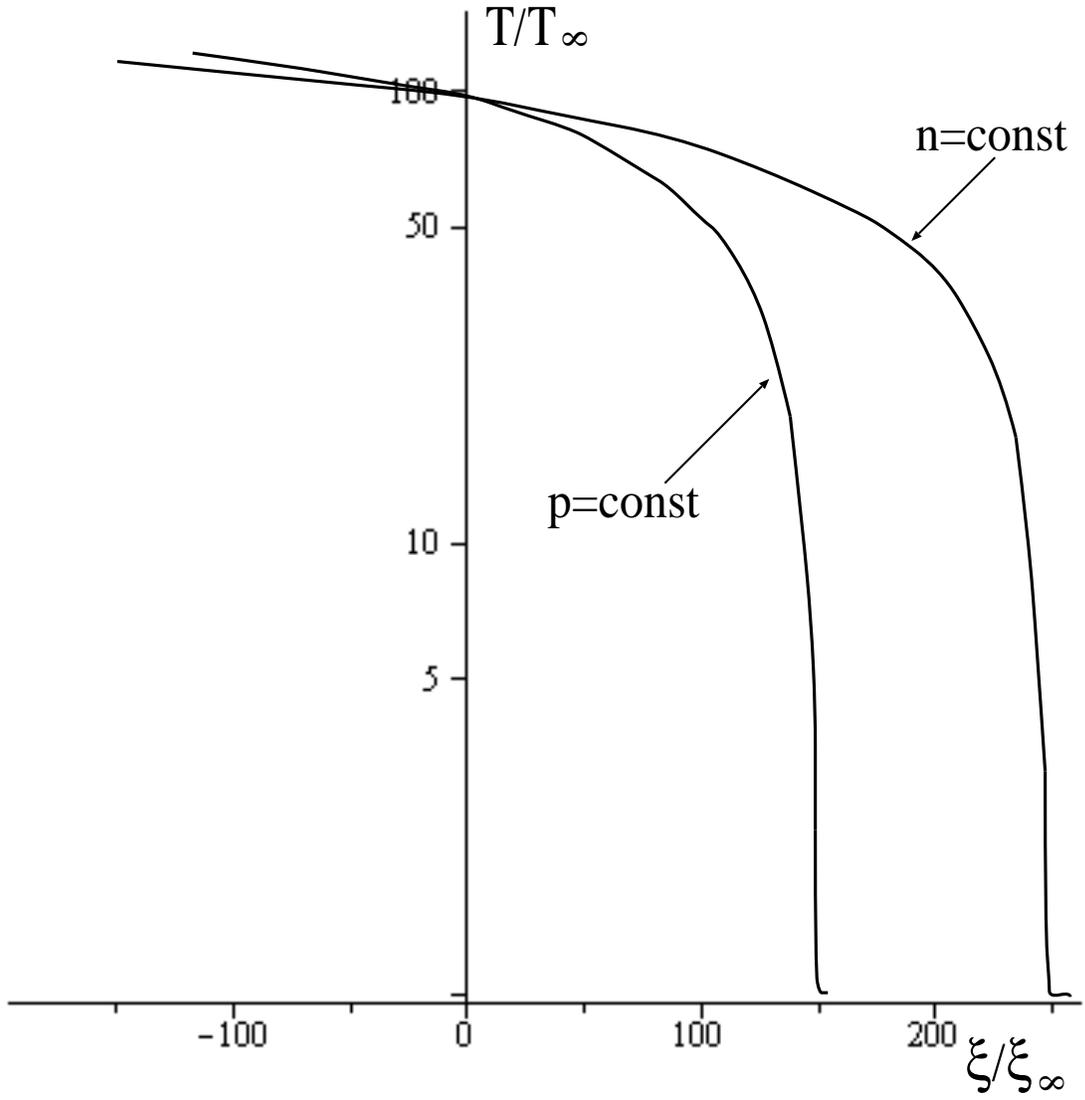}
   \caption{Distribution of the temperature along the depth $T=T(\xi)$. For the cases fast
($n=const$) and slow ($p=const$) heating.}
   \label{fig3}
\end{figure}

The dependence of the thermal flux on the temperature $F=F(T)$ for the cases $n=const$ and
$p=const$ is shown in Fig.\ref{fig2}. Here 
\begin{displaymath}
 F_\infty=\frac{\kappa(T_\infty,n_\infty)\ n_\infty}{\xi_\infty}=425\, {\rm erg/s}.
\end{displaymath}
The dependence of the temperature along the depth $T=T(\xi)$ 
for the cases $n=const$ and $p=const$
 is shown in Fig.\ref{fig3}.

%
%

\section{The width of the transition region ($\delta\xi(T)$ and $\xi_e$ comparison) }

Classical collisional heat conduction \cite{Braginskij},\cite{spitzer} is valid if two following
conditions are satisfied:

\eql{cond1}{\lambda_e<l_T=\frac{T_e}{|\nabla T_e|}}
where $\lambda_e$ is the mean free path for thermal electron collisions, $l_T$
is the characteristic scale of length on the temperature profile.

Condition \re{cond1} may be written as:
\eql{cond1-1}{\delta \xi > \xi_e,}
where $\delta \xi$ is the characteristic depth for equilibrium temperature distribution obtained
above, and $\xi_e$ is the thickness corresponding to the free path for thermal electron
collisions.

\eq{ \delta \xi (T) = \frac{ d \xi(T) }{ d \ln(T) },}
\eq{ \xi_e = n_e \lambda_e = 
     \frac{ k_{_{\rm B}}^2 T^2}{\pi e^4 \ln \Lambda}.}

\begin{figure}[ht]
   \includegraphics[scale=0.7]{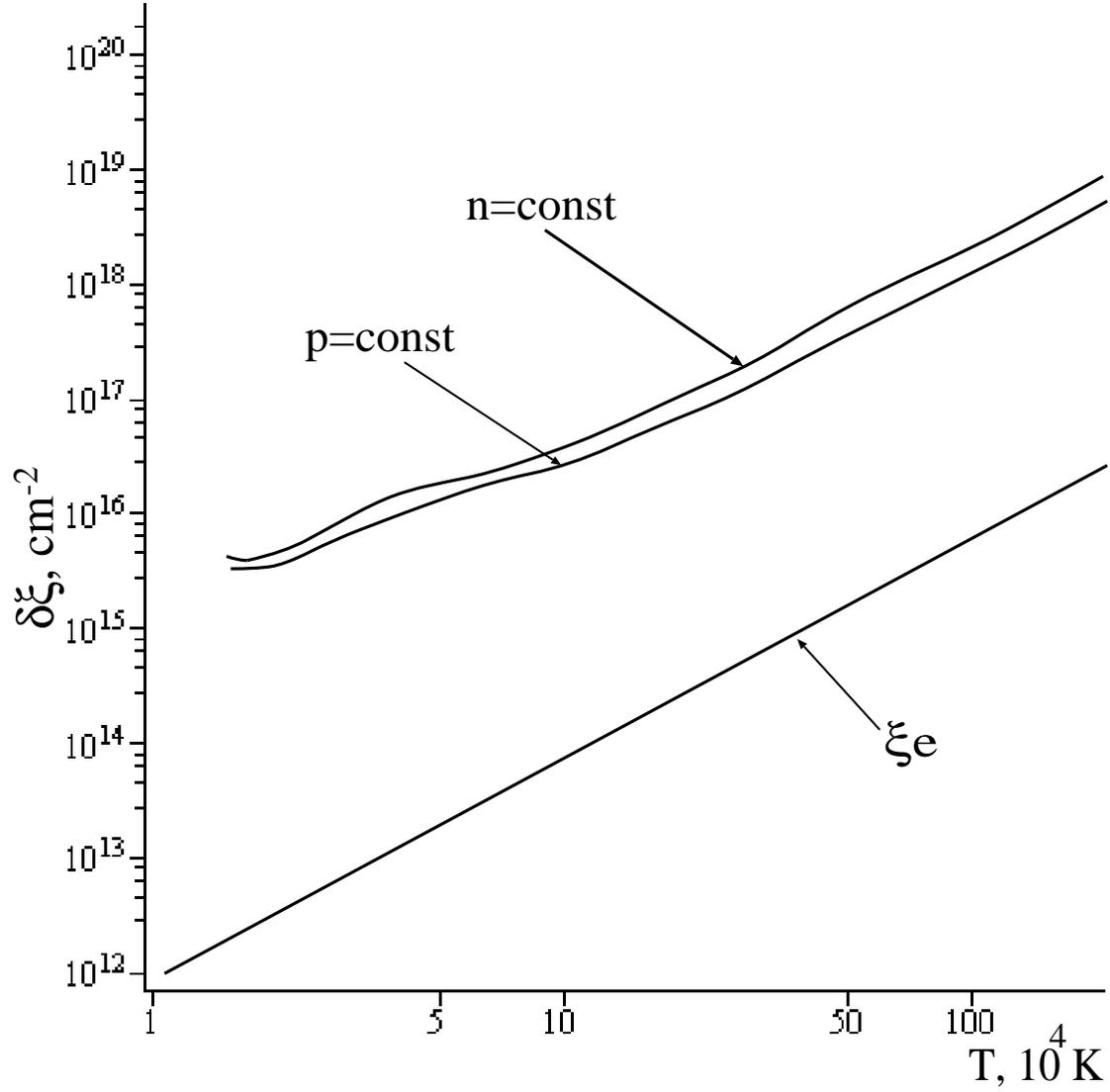}
   \caption{Characteristic values of the thickness versus the temperature for the equilibrium
temperature distriution $\delta\xi=\delta\xi(T)$ for the cases fast
($n=const$) and slow ($p=const$) heating. And the thickness corresponding to the mean free path of
thermal electrons
$\xi_e=\xi_e(T)$.}
   \label{fig4}
\end{figure}

The dependence $\delta\xi=\delta\xi(T)$ for the cases $n=const$ and $p=const$ and the dependence
$\xi_e=\xi_e(T)$ are shown in Fig.\ref{fig4}.

As we can see from Fig.\ref{fig4} $\delta\xi \gg \xi_e$, that is the characteristic thickness at
which
temperature is  changing is more then thickness corresponding to the free path for thermal
electron
collisions in 400-500 times.
For example, on the temperature $T=10^5 K$ temperature is  changing on the 35 km and the free path
for thermal
electron collisions 70 m.
 
 Thus the chromosphere-corona transition region of the solar atmosphere shall be
considered in the collisional approximation.

%
%

\section{Stability of the solution}

Linear theory of the thermal instability was constructed in monography \cite{fild} due to Field.
The uniform medium in the thermal and mechanical balance in linear theory is characterised by 3
dimensionless parameters $\alpha$, $\beta$, $\gamma$.

If the heating is such that energy gain per second to the gramm of substance is now dependent on
temperature and density, and cooling of the medium becomes formed by volume radiation energy
losses ($n^2\ L(T)$), than the parameter $\alpha$ will depend only on temperature.
And in this case $\alpha$ is the logarithmic derivative of $L(T)$

\eql{alpha}{
\alpha(T)=\frac{d\ lnL}{d\ lnT}.
}
Parameter $\beta$ characterises comparative  significance of the thermal conductivity. If
the conductivity is defined only by free electrons \re{kappa}, then

\eql{beta}{
\beta(T)=\frac{(\gamma-1)^2}{\gamma}\frac{\mu \kappa_0}{k_B^3}\sqrt{T}\ L(T)
}
That is, $\beta$ also depends only on the temperature; here $\mu$ is the effective
molecular weight (for plasma with cosmic abundance of elements $\mu \approx 1,44 m_H$).

\begin{figure}[ht]
   \includegraphics[scale=0.7]{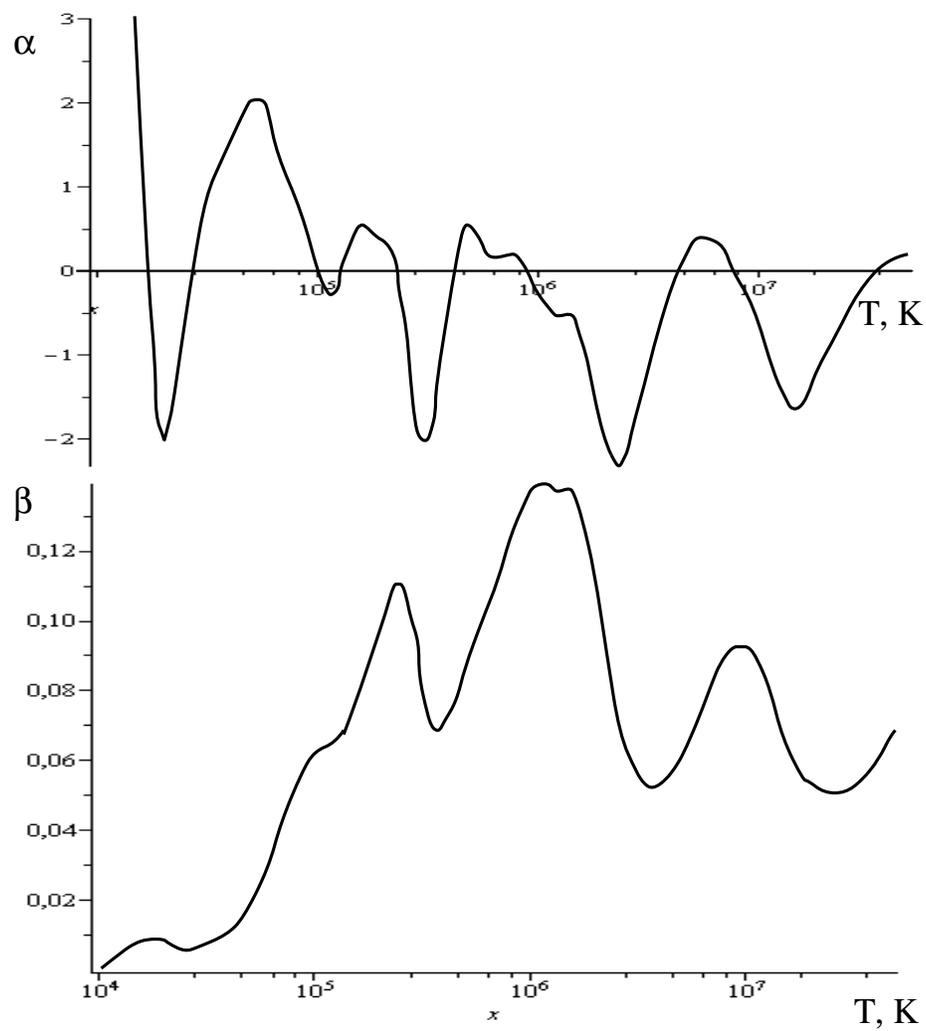}
   \caption{Distributions $\alpha=\alpha(T)$ and $\beta=\beta(T)$.}
   \label{fig5}
\end{figure}

Dependence of $\alpha$ and $\beta$ on the temperature is shown in the Fig.\ref{fig5}.
We assume $\gamma=5/3$ in the equation \re{beta}.

On figure 7 we can see regions where perturbations of the following types can be unstable: 1)
\textit{Isobaric  perturbations}, for which $p=const$. Regions where $\alpha\ <\  1$
correspond
to isobaric  perturbations. This mode is called \textit{condensation} mode of thermal instability.
2)
In regions where $\alpha < -3/2$ \textit{adiabatic} (entropy $ = const$) perturbations
are unstable. This mode of thermal instability is named \textit{wave} or 
\textit{sonic}. 3) In regions with $\ \alpha < 0$ \textit{isochoric} ($n = const$) perturbations
are unstable \cite{152}.

Let  \cite{fild},16:

\eq{
\xi_\rho=\frac{n}{k_\rho}=\frac{\gamma^{1/2}}{\gamma-1}\frac{k_B^{3/2}}{\mu^{1/2}}\frac{T^{3/2}}{
L(T)} 
,}
\eq{
\xi_T=\frac{n}{k_T}=\alpha^{-1}\ \xi_\rho
,}
\eq{
\xi_\kappa=\frac{n}{k_\kappa}=\beta\ \xi_\rho
.}

\begin{figure}[ht]
   \includegraphics[scale=0.7]{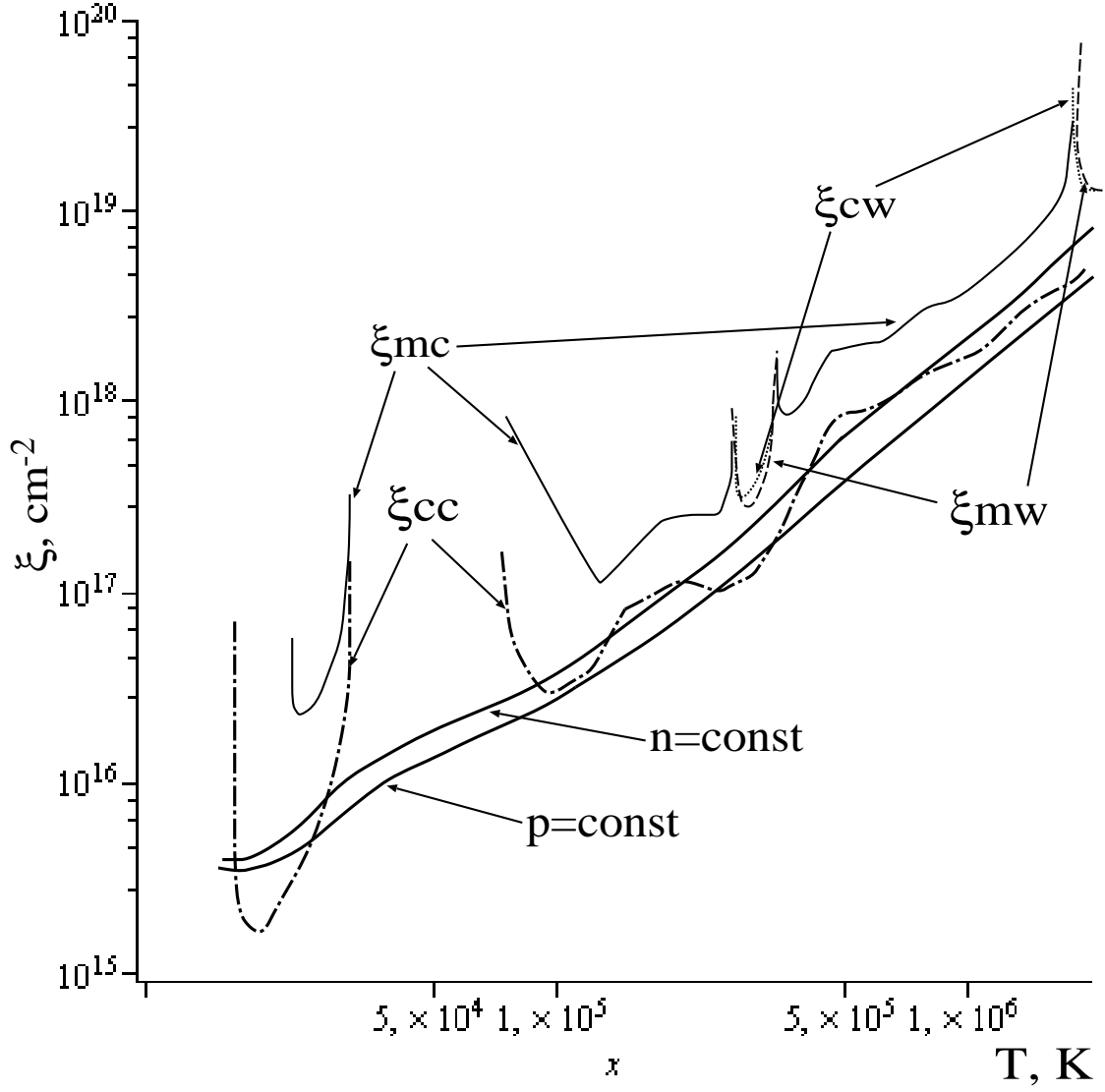}
   \caption{The scales of condensation $\xi_{cc}$ and wave
$\xi_{cw}$ perturbations,
which is stabilizate by conductivity (\re{cc},\re{cw}), and scales of perturbations with the
highest rate of growth $\xi_{mc}$,$\xi_{mw}$ (\re{mc},\re{mw})}
   \label{fig7}
\end{figure}

On the scales which are smaller than critical \cite{fild},26, thermal instability is
stabilized by condactivity:
\eql{cc}{
\xi_{cc}=\xi_\rho\ \beta^{1/2}\ (1-\alpha)^{-1/2}
,}
\eql{cw}{
\xi_{cw}=\xi_\rho\ \beta^{-1/2}\ (-\alpha-\frac{1}{\gamma-1})^{-1/2}
}
for the condensation and wave mode. These values of thickness which also depend only on
temperature, are shown in the Fig.\ref{fig7}.

The values of characteristic thickness which correspond to the biggest increase rate of
thermal instability are \cite{fild},46:

\eql{mc}{
\xi_{mc}=\left(\frac{(1-\alpha)^2}{\gamma^2}+\frac{\alpha(1-\alpha)}{\gamma}\right)^{-1/4}
\left(\xi_\rho \xi_{cc}\right)^{1/2}
,}

\eql{mw}{
\xi_{mw}=|\frac{\alpha-1}{\gamma}|^{-1/2}(\xi_\rho\ \xi_{cw})^{1/2}
}
for the condensation and wave mode, correspondingly; they are also shown in the Fig.\ref{fig7}.

The characteristic thicknesses corresponding to the stationary chromosphere heating in the cases 
$p=const$ and $n = const$ for the temperature profiles are shown in Fig.\ref{fig7}.
 At all points of distribution the balance between the thermal heating and radiative cooling has
its place.

Let us look at Fig.\ref{fig7}. If $\lambda<\lambda_{cr}$, then perturbations are smoothened out by
electron
conductivity,
if $\lambda \approx \lambda_{cr}$, then pertrubations will grow.
Note, that our solutions (in the cases $ð=const$ and $n=const$) are crossed by the curve
$\xi_{cc}$ corresponding to condensation
instability. Then, if in the places of instability there is a spontaneous change of the
temperature
profile, the profile will return to its original position.

So, we have undrestood that the temperature profile cannot be flatter than equilibrium plofile.
It also cannot be steeper, because then the same temperature will be accumulated at smaller
thickness, so the emission in this temperature range will be lower.

%
%
\section{The temperature profile emission}
The ability to emit of a certain region is named the emission measure ($ME$).

\eql{defME}{
ME=\int\limits_0^x\ n_e^2\ dl,
}
here $x$ is the length of emitting region along the line of sight, $dl$ is the interval of length.

Differential emission measure ($DME$) is the derivative of emission measure with respect to
temperature. 
\eql{defDME}{
DME=\frac{d\ ME}{d\ T}
.}

Let us rewrite $DME$ in terms of $\xi$:
\eql{DME}{
DME=\frac{d\ ME}{d\ T}=\frac{n_e^2\ dx}{dT}=n_e\ \frac{n_e\ d\ x}{d\ T}=n_e\ \frac{d\ \xi}{d\ T}
}

\begin{figure}[ht]
   \includegraphics[scale=0.7]{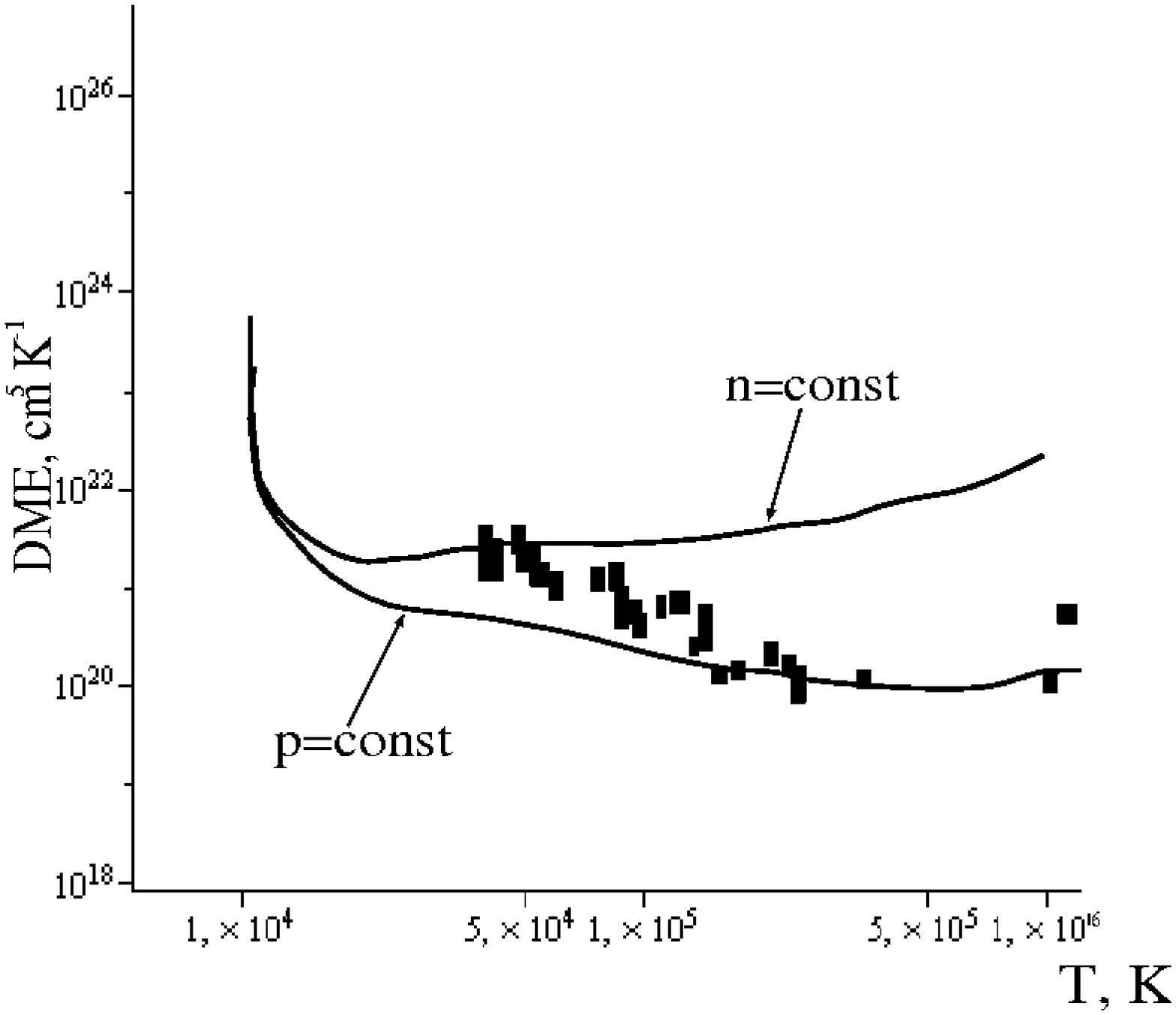}
   \caption{Dependence of the differential emission measure on the temperature $DME=DME(T)$ and
measured DME points (for different lines). }
   \label{fig8}
\end{figure}

Now, we can calculate the distribution $DME=DME(T)$ for the cases $n=const$  and $p=const$, using
the dependences  $\xi(T)$ and $n_e=n(T)$.
The result and measured DME points (for different lines) is shown in the Fig.\ref{fig8}.

%
%

\section{Conclusion}

The distribution of temperature with depth was found, assuming that the electon conductivity had
place, and at all points of distribution the balance between the
thermal heating and radiative cooling had place. Our solution is stable (see IV) and
observed UV-radiation can be explained by it (see V).

The obtained results can be used to show that temperatures are distributed in such a way
that
\textit{classical collisional} heat conduction is valid in the chromosphere-corona transition
region of the solar atmosphere, because the characteristic thickness, at which
temperature is  changing greater then thickness, corresponds to the free path for thermal electron
collisions.

%
%

\section{Acknowledgements}

O.P. thanks P. Dunin-Barkowski for useful discussions. 
This work is partly supported by RFBR grants
10-02-01315, 08-02-01033-a and by the Russian President's Grant of Support for the Scientific
Schools NSh-65290.2010.2.

\vspace*{30mm}

\vspace*{30mm}

\vspace*{30mm}

\vspace*{30mm}





\begin{thebibliography}{19}

\bibitem{besp-08} P.A. Bespalov, O.N. Savina,
{\it Astron. Lett.} {\bf 34}, N5 378 (2008).

\bibitem{besp-09} P.A. Bespalov, O.N. Savina,
{\it Astron. Lett.} {\bf 35}, N5 343 (2009).

\bibitem{Braginskij} S.I. Braginskij, in:
\textit{Leontovich M.A.(eq.) Voprosi teorii plazmy 1}, M., (1963).

\bibitem{spitzer} L. Spitzer
\textit{Physics of Fully Ionized Gases}, Wiley (1956).

\bibitem{po4ta} B.V. Somov, N.S. Dzhalilov, U. Shtaude,
 {\it Astron. Lett.} {\bf 33} N 5, 352, (2007).

\bibitem{Shm} O.P. Shmeleva, S.I. Syrovatskii,
{\it Solar Phys.} {\bf 33}, 341 (1973).

\bibitem{fild} B. Field ,
{\it Astrophys. J.} {\bf 142}, 531 (1965).

\bibitem{152} E.N. Parker,
{\it Astrophys.J.} {\bf 117}, 431 (1953).

\bibitem{Landi} E. Landi, F. Chiuderi Drago
{\it Astrophys.J.} {\bf 675}, 1629 (2008).

\bibitem{Somov} B.V. Somov,
{\it Solar Phys.} {\bf 60}, 315 (1978).

\bibitem{Somovbook} B.V. Somov,
{\it Physical Processes in Solar Flares}, Kluwer Acad. Publ.
Dordrecht, Boston (1992).

\bibitem{UFN} 	B. V. Somov, S. I. Syrovatskii, 
 {\it Uspekhi Fizicheskikh Nauk} {\bf 120}, 217 (10/1976).


\end{thebibliography}
\end{document}